\documentclass[eqsecnum,aps,epsfig]{revtex4}
\usepackage{amsmath,amssymb,amsthm,bm}
\usepackage{psfrag}
\usepackage[dvips]{graphicx}

\begin{document}
\title{Correlations between Abelian Monopoles and center vortices}
\author{Seyed~Mohsen~Hosseini Nejad$^1$ and Sedigheh~Deldar$^2$}
\affiliation{
Department of Physics, University of Tehran, P.O. Box 14395-547, Tehran 1439955961,
Iran \\
$^1$smhosseininejad@ut.ac.ir\\
 $^2$sdeldar@ut.ac.ir 
 }

\begin{abstract}
We study the correlations between center vortices and Abelian monopoles for SU($3$) gauge group. Combining fractional fluxes of monopoles, center vortex fluxes are constructed in the thick center vortex model. Calculating the potentials induced by fractional fluxes constructing the center vortex flux in a thick center vortex-like model and comparing with the potential induced by center vortices, we observe an attraction between fractional fluxes of monopoles constructing the center vortex flux. We conclude that the center vortex flux is stable, as expected. In addition, we show that adding a contribution of the monopole-antimonopole pairs in the potentials induced by center vortices ruins the Casimir scaling at intermediate regime.
\end{abstract}

\maketitle

\section{INTRODUCTION}
\label{sec:0}
The center vortices which are magnetic quantized fluxes in terms of center elements of the gauge group are line-like objects in three dimensions or surface-like in four dimensions. The random fluctuations of the number of center vortices linked to the Wilson loop leads to quark confinement \cite{DelDebbio:1996mh,Langfeld:1997jx,Kovacs:1998xm,Alexandrou:1999iy,Engelhardt:1999fd,Engelhardt:1999wr,Engelhardt:2003wm,DelDebbio:1997}. On the other hand, the condensation of the Abelian monopoles leads to the quark confinement in the dual superconductor scenario \cite{Suzuki1993,Hooft1981}. The presence of both center vortices and Abelian monopoles in the vacuum seems to be an indication of some kind of correlations between these objects. According to Monte Carlo simulations, a center vortex in SU($2$) gauge group upon Abelian projection would appear as a monopole-vortex chain \cite{Del Debbio1998,Ambjorn2000} and also for the SU($3$) case, there are correlations between center vortices and monopole sources \cite{Stack2002}. For general case of SU($N$) gauge group, $N$ center vortex fluxes meet at a monopole-like center and construct monopole-vortex junctions \cite{Cornwall1977,Cornwall1998,Chernodub2008,HD2016}. 

In this article, the monopole-vortex configurations are studied in SU($3$) gauge group. The motivation is to study
the correlation between the center vortices and monopoles. Even though lattice simulations show that both configurations lead to confinement and the existence of the monopole-vortex chains are supported by numerical simulations \cite{Del Debbio1998,Ambjorn2000,Stack2002}, however studying the details of the monopole or vortex fluxes are not possible by lattice calculations. It is much harder for SU($N>2$) gauge theory which has more than one monopole type. In this work, we first construct the center vortex flux of SU($3$) gauge theory with combining fractional fluxes of monopoles obtained from Abelian gauge fixing. Some fractional flux configurations are constructed and combinations of them produce monopole-vortex configurations.
Then, using a thick center vortex-like model \cite{Fabe1998}, we study the potentials induced by monopole fractional flux configurations which are line-like similar to center vortices.
Comparing the potential induced by fractional fluxes of monopoles constructing the center vortex flux in a thick center vortex-like model with the one induced by center vortices, we observe an attractive energy between fractional fluxes of monopoles constructing the center vortex flux. Therefore, we conclude that center vortex fluxes are stable configurations, as expected. 

Moreover, the static potentials for some representations are calculated using center vortices. The results agree the Casimir scaling at intermediate distances. The effect of line-like monopole-vortex configurations on the Wilson loop is the same as the effect of center vortices on the loop. Then, we assume the existence of monopole-antimonopole pairs in the vacuum in addition to center vortices. These pairs carry Abelian flux of monopoles. We show that if the monopole-antimonopole pairs as Abelian configurations were available, they would ruin the Casimir scaling effect at intermediate regimes. According to the lattice results, the Abelian U($1$)$^2$ subgroup of SU($3$) cannot account for Casimir scaling, while the string  tension of an Abelian-projected loop agree with the $N$-ality at asymptotic distances \cite{Greensite2007,Fabe1996}.
 It seems that in Abelian theories there are the configurations similar to monopole-antimonopole pairs with Abelian fluxes which ruin the Casimir scaling effect. 

 In section \ref{sec:1}, the formation of Abelian monopoles appeared by the Abelian gauge fixing method is recalled for SU($3$) gauge group. The correlations between the Abelian monopoles and center vortices reported in lattice gauge theories and other phenomenological models, are described in section \ref{sec:2}. Then, the thick center vortex model is briefly reviewed in section \ref{sec:3}. The center vortices as a combination of fractional fluxes of Abelian monopoles in SU($3$) gauge theory is investigated in section \ref{sec:4}. Stability of center vortex configurations constructed from the fractional fluxes of Abelian monopoles is discussed in section \ref{sec:5}. We study the potential ratios induced by center vortex and monopole fluxes and compare them with Casimir ratios in section \ref{sec:7}. Finally, a summary of our study is given in section~\ref{sec:8}.

\section{Magnetic monopole charges}
\label{sec:1}

In the non Abelian gauge theories, Magnetic monopoles are appeared by Abelian gauge fixing. In general, one can do a gauge transformation to reduce a non Abelian gauge theory into an Abelian gauge theory by diagonalizing a specific field of the theory. However, the gluon field is not a good candidate; because only one of the four components of the gluon field can be aligned simultaneously. Therefore, a scalar field can be used for fixing an Abelian gauge. After Abelian gauge fixing (see Ref. \cite{Ripka}), the gauge field under a gauge transformation which diagonalizes the scalar field obtains singularity in the vicinity of specific points in
space where Abelian gauge fixing becomes undetermined and magnetic monopoles are formed. For $SU\left( 3\right) $ gauge theory which has two diagonal generators $\mathcal{H}_3$ and $\mathcal{H}_8$, the topological defects of Abelian gauge fixing are sources of magnetic monopoles with magnetic charges equal to: 
\begin{equation}
\label{g}
g_i=\frac{4\pi }{e}\left( w_{i}\cdot \mathcal{H}\right),
\end{equation}
where $\mathcal{H}$ is the vector $\left( \mathcal{H}_3,\mathcal{H}_8\right) $, $e$ is the color electric charge and $w_{i=1,2,3}$ are the root vectors of group as the following
\begin{equation}
w_1=\left( 1,0\right), \;\;\;\;\;\;w_2=\left( -\frac 12,-\frac{\sqrt{3}}%
2\right), \;\;\;\;\;\;w_3=\left( -\frac 12,\frac{\sqrt{3}}2\right).
\label{apsu:weight}
\end{equation}

\section{Magnetic monopoles and center vortices}
\label{sec:2}

\begin{figure}
\begin{center}
%\vspace{90pt}
\resizebox{0.5\textwidth}{!}{
\includegraphics{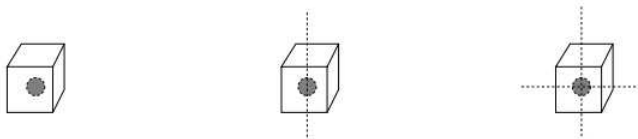}}
%\vspace{-40pt}
\caption{\label{1}
Monopoles passed by center vortex line. Almost all monopoles (about 93\%) are passed through one center vortex line (middle panel). The small fractions of monopoles either are not passed through center vortex lines
   at all (about 3\%)(left panel), or are passed by more than one line (about 4\%)(right panel) \cite{Del Debbio1998}.}
\end{center}
\end{figure}

Both the monopole and center vortex mechanisms of the quark confinement are supported by numerical
simulations. Therefore, the fluxes of the monopoles and center vortices are expected to be interrelated. According to the lattice Monte Carlo simulations in SU($2$) gauge theory, after Abelian projection almost all monopoles are sitting on top of the center vortices \cite{Del Debbio1998,Ambjorn2000} as shown in Fig. \ref{1}. Almost all static monopoles located in the unit cubes on the lattice are passed by a center vortex line. The small fractions of monopoles are not either
passed through any center vortex line at all, or are passed by more than one line. After Abelian projection, about $61\%$ of vortex  lines  have  no  monopoles  at  all  on  them and  $31\%$  have a monopole+antimonopole. The remaining  $8\%$ of vortex  lines  contain an  even  number  of 
monopoles+antimonopoles. 
On the other hand, lattice results have shown that center vortices meet monopole sources in SU($3$) gauge theory \cite{Stack2002}. 

As a result, numerical simulations have indicated the correlations between center vortices and Abelian monopoles. In Ref. \cite{HD2016}, we have studied these correlations for SU($2$) gauge theory in a thick center vortex-like model 
where monopole-antimonopole pairs which are line-like configurations split into two center vortex fluxes and monopole-vortex chains appear in the model.

\begin{figure}[]
\vspace{100pt}
\centering
a)\includegraphics[width=0.20\columnwidth]{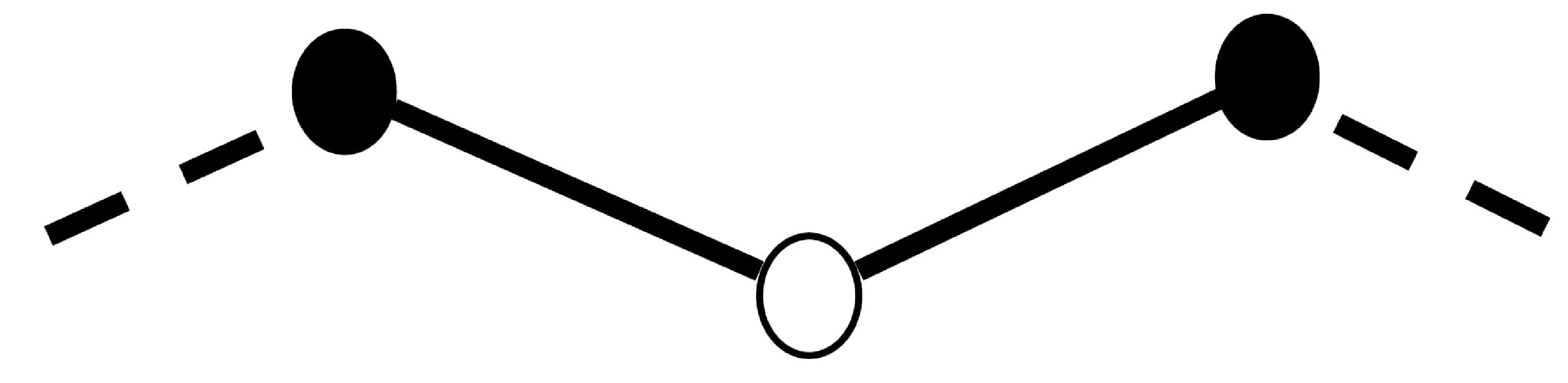}
b)\includegraphics[width=0.2\columnwidth]{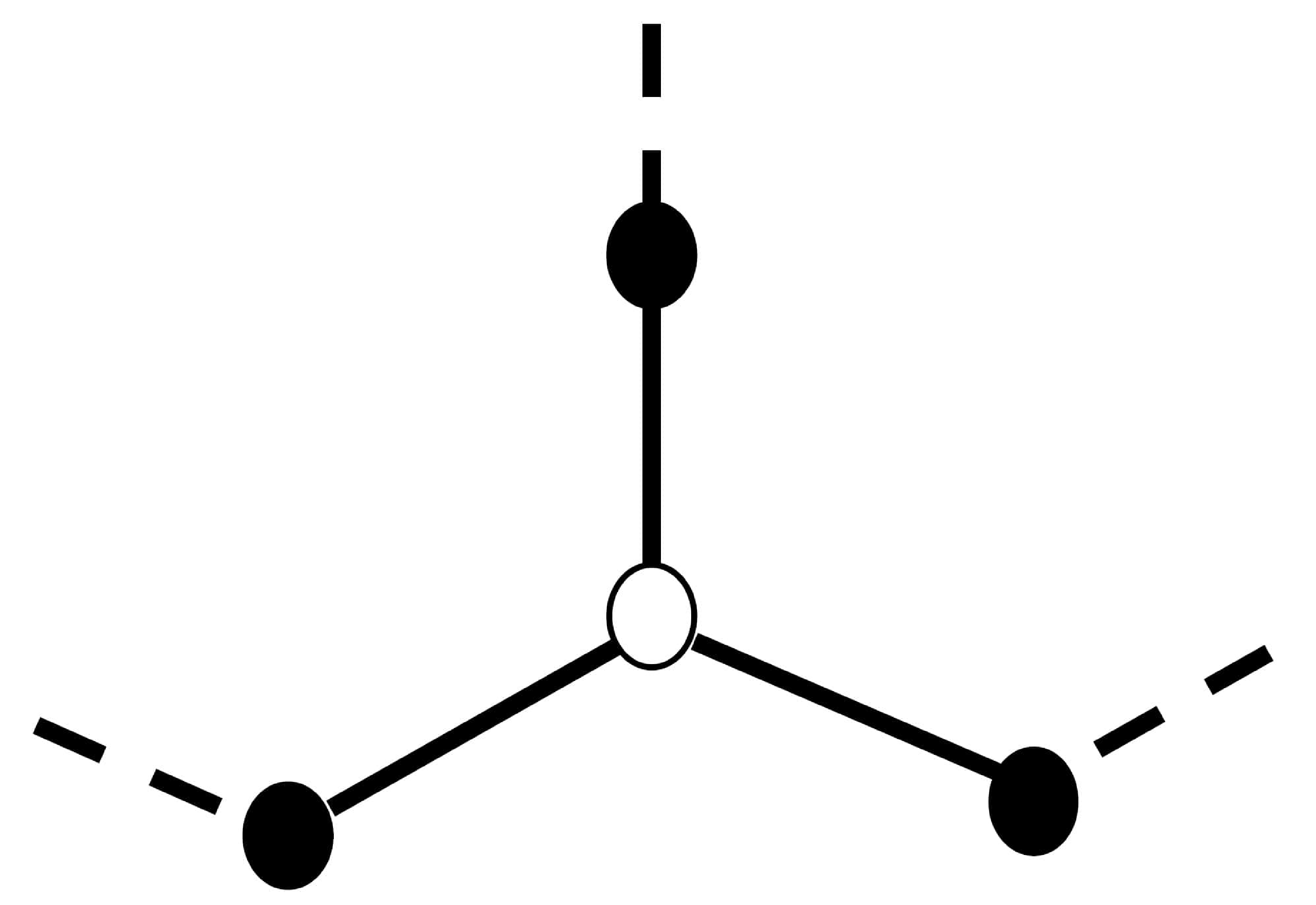}
\caption{a) The monopole-vortex 
chains in SU($2$) gauge theory b)  the monopole-vortex nets in SU($3$) gauge theory \cite{Ambjorn2000,Cornwall1998,Chernodub2008}.}\label{2}
\end{figure}

 In addition, the monopole-vortex junctions are studied in Ref. \cite{Cornwall1977} where they are called as nexuses. Some solutions of the equations of motion coming from the low-energy effective energy functional 
$E$ of QCD \cite{Cornwall1998} are studied where several center vortices meet at a monopole-like center (nexus). In SU($2$) gauge theory, two center vortices meet at the nexus and therefore monopole-vortex chains appear in the theory while in SU($3$) gauge theory three center vortices meet at the nexus and monopole-vortex nets are observed as shown in Fig. \ref{2}. In SU($N$) gauge group, $N$ center vortices meet at the nexus.

\section{Thick center vortex model}
\label{sec:3}
 In this model, the vacuum is dominated by line-like center vortices which have a thickness. It is assumed that the effect of a center vortex on a Wilson loop is to  multiply  the 
loop by a group element \cite{Fabe1998}
\begin{equation}
 \label{group}
{G}(\alpha_C^{(n)})=\mathcal{G}_r(\alpha_C^{(n)})\; {I}_{d_r}=\frac{1}{d_r}Tr\left(\exp\left[i\vec{\alpha}_C^{(n)}\cdot\vec{\mathcal{H}}\right]\right)\; {I}_{d_r},
\end{equation}
where $\{\mathcal{H}_i~i=1,..,N-1\}$ are Cartan generators of SU($N$) gauge group. $\mathcal{G}_r(\alpha_C^{(n)})$ is the group factor and $d_r$ is the dimension of the representation $r$. The flux profile $\vec{\alpha}_C^{(n)}(x)$ depends on the position $x$ of the center vortex relative to the Wilson loop, and $n$ indicates the center vortex 
type.

If the center vortex locates entirely within the Wilson loop, then 
  \begin{equation}
   \label{z}
\exp\left[i\vec{\alpha}_C^{(n)}(x)\cdot\vec{\mathcal{H}}\right]=z_n I_N,
\end{equation}
where $z_n=e^{{2\pi i n}/{N}}$ and $I_N$ is the $N \times N$ unit matrix.
 
The quark potential induced by center vortices in representation $r$ (see Ref. \cite{Fabe1998}) is obtained as the following 

\begin{equation}
\label{potential}
V_r(R) = -\sum_{x}\ln\left\{ 1 - \sum^{N-1}_{n=1} f_{n}
(1 - {\mathrm {Re}}\mathcal{G}_{r} [\vec{\alpha}^{(n)}_{C}(x)])\right\},
\end{equation} 
here an $n$th center vortex pierces any given plaquette with the probability $f_n$. 
Assuming intermediate Wilson loops associated with small $\alpha$ and also $f_n<<1$, the potential induced by center vortices agrees with the Casimir scaling for intermediate distances. Therefore, for observing the property of Casimir scaling in the potential, the probability $f_n$ should be far smaller than $1$. 
An appropriate ansatz for the vortex profile $\vec{\alpha}_C^{(n)}(x)$
which respects linearity and Casimir scaling for the intermediate regime of the quark potential is \cite{Fabe1998}:
\begin{equation}
\label{alpha}
\alpha_i^{(n)}(x)=\frac{\alpha_i^{n(max)}}{2}[1-\tanh(ay(x)+\frac{b}{R})], ~~~~~~~~y(x) = \left\{ \begin{array}{cl}
                     x-R & \mbox{for~} |R-x| \le |x| \cr
                     -x  & \mbox{for~} |R-x| > |x|
                   \end{array} \right.
\end{equation}
where $a, b$ are free parameters of the model and $\alpha_i^{n(max)}$ is the maximum value of the angle. 
We recall that, the Casimir scaling effect is not observed at intermediate distances for any
choice of the free parameters $a, b$. However, it is found for a large range of the parameters. For example, the extent of Casimir scaling region at intermediate distances can be rescaled by any factor $F$ by setting $a \rightarrow a/F,~b \rightarrow bF$. For the ansatz given in Eq. (\ref {alpha}), the vortex thickness would be of the order of ${1}/{a}$. Therefore, $F>1$ increases the vortex thickness and the Casimir scaling region while $F<1$, decreases these quantities \cite{Fabe1998,HD2014,HD2015,HD2016}. 

Some other ansatz for the vortex profile $\vec{\alpha}_C^{(n)}(x)$ which results to appropriate potentials are discussed in Ref. \cite{Deldar2001}. The same physical results are obtained for appropriate physical vortex profiles as discussed in that reference.

In the next section, using the fractional fluxes of magnetic monopoles, we obtain center vortex fluxes.

\section{ Combining fractional fluxes of Abelian monopoles and center vortices}
\label{sec:4}

 According to section \ref{sec:2}, monopoles and center vortices correlate to each other and therefore one may expect that the fluxes of the monopoles and center vortices are interrelated. Now, we construct the center vortex flux of SU($3$) gauge theory with combining fractional fluxes of monopoles using the thick center vortex model \cite{Fabe1998}.
 For SU($3$) gauge group, there is a non trivial center element $z_1=e^{\frac{2\pi i}{3}}$ and the quantized flux which the center vortex carries is equal to $\phi_v={\frac{2\pi}{3}}$. Using Eq. (\ref{z}), the maximum values of the angles corresponding to the Cartan generators $\mathcal{H}_{3}$ and $\mathcal{H}_{8}$ for the fundamental representation are equal to zero and $\frac{4\pi}{\sqrt{3}}$, respectively. Therefore, we get 
\begin{equation}
\label{center3}
\exp\left[i\frac{4\pi}{\sqrt{3}} \mathcal{H}_8\right]=z_1 I,
\end{equation}
if the center vortex locates entirely within the Wilson loop.
On the other hand, the total flux of a magnetic monopole crossing the closed surface $S$ surrounding the monopole is equal to \cite{Chatterjeea2014}: 
\begin{equation}
\label{ce}
{\Phi}_m=\int_S\vec{B}.d\vec{s}=g,
\end{equation}
where $g$ is the magnetic charge of the monopole. In SU($3$) gauge group, magnetic charges given in Eq. (\ref{g}) satisfy the constraint
\begin{equation}
\label{constraint}
g_1+g_2+g_3=0.
\end{equation} 
As a result, the number of independent magnetic charges reduces to $2$. Using magnetic charges in Eq. (\ref{g}), the Cartan generator $\mathcal{H}_{8}$ is obtained versus $g_3$ and $g_2$ : $\mathcal{H}_{8}=\frac{e}{4\pi\sqrt{3}}(g_3-g_2)$. Substituting $\mathcal{H}_8$ in Eq. ({\ref{center3}) gives a suggestion for relations between center vortices and monopoles.
\begin{equation}
\label{cen1}
\exp\left[ie(\frac{g_3}{3}-\frac{g_2}{3})\right]=z_1 I.
\end{equation}
Therefore, according to Eq. (\ref{cen1}) the effect of a center vortex on the Wilson loop is equivalent to the effect of an Abelian configuration corresponding to one third of the matrix flux $g_3-g_2$ on the loop. In general, the contribution
of an Abelian field configuration to the Wilson loop related to $q$ units of the electric charge is $W=e^{iq{\Phi}}$ where for the fundamental representation $q=1$ \cite{Chernodub2005}. Back to Eq. (\ref{group}), the contribution of the Abelian configuration, the left hand side of Eq. (\ref{cen1}), to the Wilson loop is as the following
   \begin{equation}
\label{c8}
W=\mathcal{G}_f=\frac{1}{d_f}\mathrm{Tr}\left(\exp\left[ie(\frac{g_3}{3}-\frac{g_2}{3})\right]\right)=\frac{1}{3}\mathrm{Tr}\left(\begin{array}{ccc} e^{i\frac{2\pi}{3}} & 0 &0\\0 &e^{i\frac{2\pi}{3}}&0\\ 0 &0 &e^{-i\frac{4\pi}{3}}
\end{array} \right)=e^\frac{i2\pi}{3},
\end{equation} 
which is obtained to be equal to the right hand side of Eq. (\ref{cen1}) that is the vortex contribution.
Therefore, the flux of the Abelian configuration corresponding to one third of the matrix flux $g_3-g_2$ is $\frac{2\pi}{3}$. 
In other words, the center vortex carries one third of the total monopole flux $g_3$ plus one third of the total monopole flux $g_2$ pointing in opposite direction. This result is obtained where the center vortex is completely located within the Wilson loop and is independent of the ansatz of the angle and therefore the model. 

As a result, some monopole fractional flux configurations which may appear in the SU($3$) monopole vacuum are plotted  
 in Fig. (\ref{fig3}). The fractional flux lines with one arrow carry one third of the total flux of a monopole and the ones with two arrows carry two third of the total flux of a monopole.
 In general, for SU($3$) gauge group, three flux lines emerge from each monopole. Each flux line starts from a monopole and ends to an antimonopole. In Fig. (\ref{fig3}) (left-up), three fluxes emerge from $g_3$ type and enter to the antimonopole of $g_3$. The same is drawn for $g_2$ type monopole in the bottom. Fig. (\ref{fig3})(left up and down) are closed chains which are not helpful in calculating the potentials between quarks as we explain later. Monopole-antimonopole may sit in a line as shown in Fig. (\ref{fig3})(right), as observed in lattice gauge theory, as well. The arrows emerge from monopoles and enter to the antimonopoles. There are two plots for two monopole types of SU($3$) gauge group. 
SU($3$) monopole charges satisfy the Dirac quantization condition $eg=2n\pi$, $\exp\left[\pm ieg\right]=1$. Therefore, two third of the total monopole flux on the Wilson loop may be regarded the same as one third of the total monopole flux pointing in opposite direction
\begin{equation}
\label{cenn}
\exp\left[ie\frac{2g_n}{3}\right]=\exp\left[ie\frac{2g_n}{3}-ieg_n\right]=\exp\left[ie\frac{-g_n}{3}\right],
\end{equation}
where $g_n$ are SU($3$) monopole charges in Eq. (\ref{g}). Therefore, using Dirac quantization condition a fractional flux line with two arrows as shown in Fig. \ref{fig3} may be regarded the same as a line with one arrow pointing in opposite direction.
Fig. \ref{fig3}(right) is re-plotted in Fig. \ref{fig4} using Dirac quantization condition. The fractional flux lines with two arrows in Fig. \ref{fig3}(right) are drawn with one arrow in Fig. \ref{fig4} with opposite arrow. This orientation is appropriate for constructing vortices as discussed later.  

Back to Eq. (\ref{cen1}), the monopole fractional fluxes have the same effect on the Wilson loop as the center vortex when they are located completely within the Wilson loop. Therefore combining the monopole fractional flux configurations given in Figs. \ref{fig3} and \ref{fig4} may produce monopole-vortex configurations as shown in Fig. \ref{fig5}. The lines in the configurations carry the center vortex fluxes which are a combination of a fractional flux line corresponding to one third of the total flux of $g_3$ monopole and a line corresponding to one third of the total flux of $g_2$ monopole pointing in opposite direction.
\begin{figure}
\includegraphics[scale=0.05]{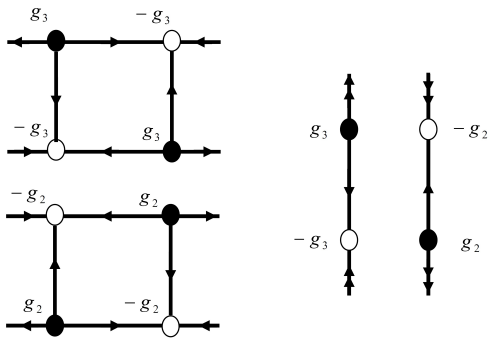}
\caption{Some monopole fractional flux configurations for SU($3$) gauge theory. Black and white circles indicate monopoles and antimonopoles, respectively. The fractional flux lines with one arrow carry one third of the total flux of a monopole and the ones with two arrows carry two third of the total flux of a monopole. }
\label{fig3}
\end{figure}
 \begin{figure}
\includegraphics[scale=0.055]{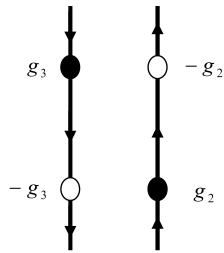}
\caption{Same as the right panel of Fig. \ref{fig3} but using Dirac quantization condition. The fractional flux lines with two arrows of figure 3 (right)  may be regarded the same as a line with one arrow pointing in opposite direction in this figure.}
\label{fig4}
\end{figure}
\begin{figure}
\vspace{50pt}
\includegraphics[scale=0.055]{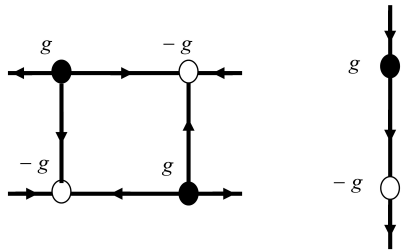}
\caption{Some monopole-vortex configurations in SU($3$) gauge theory. The lines in the configurations carry the center vortex fluxes and the magnetic charge $g$ is equal to $g_3-g_2$.}
\label{fig5}
\end{figure}

 Now we investigate stability of center vortices, by comparing the potentials induced by monopole fractional fluxes in a thick center vortex-like model and the one induced by center vortices. But before that we explain thick center vortex-like model where the effect of some monopole fractional flux configurations which are line-like similar to center vortices is studied on the Wilson loops.

\section{ fluxes of Abelian monopoles on the vacuum}
\label{sec:5}

In the thick center vortex model, the center vortices are line-like and the Wilson loop is characterized  
just by the width $R$, the distance between quark and antiquark.  Now we study the configurations with the fluxes of monopoles which are line-like similar to center vortices in a thick center vortex-like model. This model is the same as thick center vortex model but the effect of the fluxes of monopoles on the Wilson loop is investigated. 

We assume the magnetic fields between monopole and antimonopole are localized in a tube with a thickness which are line-like similar to center vortices. The effect of locating a monopole-antimonopole configuration within the Wilson loop is represented by insertion of a phase $e^{ ie\int_S\vec{B}.d\vec{s}}$ \cite{Chernodub2005} in the link product of the Wilson loop where $e$ is the color electric charge and $\int_S\vec{B}.d\vec{s}$ is the total magnetic flux of the monopole which is equal to the magnetic charge of monopole according to  Eq. (\ref{ce}). In other words, the effect of a monopole-antimonopole configuration on a Wilson loop is to multiply the loop 
by a group element, the same as the one in Eq. (\ref{group}). If a monopole-antimonopole configuration locates entirely within the Wilson loop, then
\begin{equation}
\label{center}
\exp\left[i\vec{\alpha}^{(n)}\cdot\vec{\mathcal{H}}\right]=e^{ ieg},
\end{equation}
where in SU($3$) gauge group the magnetic charge of monopole $g$ satisfying the Dirac quantization condition $eg=2n\pi$ is given in Eq. (\ref{g}). 

Among the monopole fractional flux configurations plotted in the previous section, the configurations shown in Fig. \ref{fig4} are line-like and affect the Wilson loop while the effect of a fractional flux inside a configuration plotted in the left panel of Fig. \ref{fig3}, is eliminated by another fractional flux pointing in opposite direction and has no effect on the Wilson loop. 
If a monopole fractional flux configuration in Fig. \ref{fig4} locates entirely within the Wilson loop, then
\begin{equation}
\label{fraction}
\exp\left[i\vec{\alpha}^{(n)}\cdot\vec{\mathcal{H}}\right]=e^{ ie\frac{g_n}{3}},
\end{equation}
where the fractional flux is one third of the total flux of magnetic monopole $g_n (n=2,3)$. The potential induced by the line-like monopole fluxes is obtained by a similar method used for calculating the potential induced by the center vortices. Therefore,  the potential induced by the line-like monopole fluxes is the same as Eq. (\ref{potential}) but the index $n$ corresponds to the line-like monopole fluxes.

\section{ Center vortices as stable composites of monopole fluxes}
\label{sec:6}
 In the previous sections, we have shown that combining one third of the total flux of $g_3$ monopole and one third of the total flux of $g_2$ monopole pointing in the opposite direction leads to the center vortex flux. To understand the interaction between these fractional fluxes constructing the center vortex flux, we study the potentials induced by
fractional fluxes lines of the monopoles and center vortices. In the thick center vortex model, the string tension or the potential energy between static quark-antiquark for a representation depends only on the flux profile ${\alpha}_C^n$ of the center vortices and the piercing probability $f_n$ of any given plaquette by center vortices. Increasing these quantities increases the magnetic energy of the vacuum and therefore the potential energy between static quark-antiquark. As a result, the potential energy between static quark-antiquark changes because of changing the magnetic energy of the vacuum. On the other hand, the interaction energy between center vortices is the magnetic energy type and changing this energy leads to the change of the potential energy between static quark-antiquark \cite{HD2015}. Now, we compare the potentials induced by center vortices and fractional fluxes lines of the monopoles constructing the center vortex fluxes. 
 
 Using Eqs. (\ref{fraction}) and (\ref{potential}), the potential induced by fractional flux lines of the monopoles (see Fig. \ref{fig4}) where combinations of them produce the center 
vortex fluxes is as the following:
\begin{equation}
\label{potential2}
V_f(R) = -\sum_{x}\ln\left\{ 1 - \sum^{3}_{n=2} f_{n}
(1 - {\mathrm {Re}}\mathcal{G}_{f} [\vec{\alpha}^n_{C}(x)])\right\},
\end{equation}
where $f$ denotes the fundamental representation and the fractional flux configurations enumerated by the value $n=2,3$ correspond to one third of the total flux of magnetic monopole $g_n$ in Eq. (\ref{g}). When the fractional flux lines of the monopoles locate entirely within the Wilson loop, we use Eq. (\ref{fraction}). Therefore, the maximum values of the angles for one third of the total flux of magnetic monopole $g_2$ corresponding to the Cartan generators $\mathcal{H}_{3}$ and $\mathcal{H}_{8}$ for the fundamental representation are equal to $\frac{2\pi}{3}$ and $-\frac{2\pi}{\sqrt{3}}$ while these values for one third of the total flux of magnetic monopole $g_3$ are $\frac{2\pi}{3}$ and $\frac{2\pi}{\sqrt{3}}$, respectively.

On the other hand, the potential induced by $z_1$ center vortices is 
\begin{equation}
\label{potential3}
V_f(R) = -\sum_{x}\ln\left\{ 1 -  f_{1}
(1 - {\mathrm {Re}}\mathcal{G}_{f} [\vec{\alpha}^1_{C}(x)])\right\}.
\end{equation}
Using Eq. (\ref{z}) for SU($3$) gauge group where $z_1=e^{\frac{2\pi i}{3}}$, the maximum values of the angles corresponding to the Cartan generators $\mathcal{H}_{3}$ and $\mathcal{H}_{8}$ for the fundamental representation are equal to zero and $\frac{4\pi}{\sqrt{3}}$, respectively.
Figure \ref{fig6} shows the static potential of the fundamental representation induced by fractional fluxes of the monopoles compared with the one induced by the center vortices.
\begin{figure}
\begin{center}
%\vspace{90pt}
\resizebox{0.45\textwidth}{!}{
\includegraphics{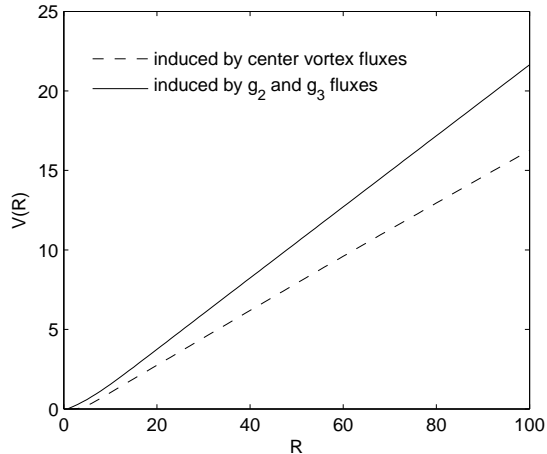}}
%\vspace{-40pt}
\caption{\label{fig6}The potential energy induced by  monopole fractional fluxes corresponding to one third of $g_3$ monopole fluxes and one third of $g_2$ monopole fluxes where combination of them produce the center 
vortex fluxes and the one obtained by the center vortices. The extra negative energy of potential induced by center vortex compared with the one induced by fractional fluxes of monopoles shows that the fractional flux lines, making center vortex fluxes, attract each other and make stable configurations. The free parameters $f_n$, $a$ and $b$ are chosen to be $0.1$, $0.05$ and $4$, respectively.}
\end{center}
\end{figure}
The free parameters $f_n$, $a$ and $b$ in Eqs. (\ref{potential2}) and (\ref{potential3}) are chosen to be $0.1$, $0.05$ and $4$, respectively. As shown in Fig. \ref{fig6}, the behavior of the potential induced by fractional fluxes of monopoles is similar to the one induced by center vortices and they are both linear. The potential energy induced by fractional fluxes, corresponding to one third of $g_3$ monopole fluxes and one third of $g_2$ monopole fluxes, is larger than the potential energy induced by the center vortices at intermediate and large distances. At large distances, this result is independent of the ansatz of the angle and therefore the model. According to the model, the string tension of fundamental
representation sources are the same at intermediate and large distances \cite{Fabe1998}. Therefore, the ultimate physical result is not affected at intermediate distances by changing the vortex profile.

Both potentials are obtained by an ensemble of the monopole fractional flux lines plotted in Fig. \ref{fig4}. However, there is a difference between the two potentials. In calculating the potential energy induced by fractional fluxes, the two monopole fractional flux lines plotted in Fig. \ref{fig4} are assumed to be independent and without any interaction while in the potential energy induced by center vortices, the presumed  monopole fractional flux lines plotted in Fig. \ref{fig4} appear in the vacuum simultaneously and interact each other. Therefore, we expect some change of magnetic energy as an interaction between the monopole fractional flux lines constructing the center vortex flux.
As mentioned above, changing the magnetic energy of the vacuum changes the quark potential. The extra negative energy of the potential induced by center vortices compared with the one induced by fractional fluxes of monopoles shows that the fractional flux lines (see Fig. \ref{fig4}), attract each other. As a result, the center vortex fluxes are stable configurations due to the attraction between the fractional fluxes. 

On the other hand, in Ref. \cite{Reinhardt}, Reinhardt et al. explained that the monopole-antimonopole flux line in SU($2$) gauge group is split into two equal portions of center vortex fluxes and construct a monopole-vortex chain, as shown by the lattice simulations. Therefore, there are repulsions between two magnetic vortex fluxes with the same flux orientations. In other words, there are attractions between two magnetic fluxes with opposite flux orientations. Therefore, It is also confirmed by other theories that the center vortex flux is constructed from the fractional flux lines with opposite flux orientations attracting each other. 

\section{ Casimir scaling and Monopole fluxes}
\label{sec:7}
Based on numerical  simulations, the string tensions of the static color source potentials at intermediate distances, from the onset of confinement  to  the  onset  of  charge  screening is expected to be linearly rising and agree with the Casimir scaling. At this regime, the string tension for a color source in representation $r$ is approximately proportional to the eigenvalue of the quadratic Casimir operator $C_r$ of the representation, $\it{i.e.}$
 \begin{equation}
\label{casimir}
\sigma_r=\frac{C_r}{C_f}\sigma_f,
\end{equation}
where $f$ denotes the fundamental representation. According to Ref. \cite{DelDebbio:1997}, since center vortices of monopole-vortex configurations are condensed, the monopoles of them are condensed, as well. The effect of line-like monopole-vortex configurations, as shown for SU($3$) gauge group in right panel of Fig. \ref{fig5}, on the Wilson loop is the same as the effect of center vortices on the loop. In SU($3$) case, there are two nontrivial
center elements $z_1=e^{\frac{2\pi i}{3}}$ and $z_2=e^{\frac{4\pi i}{3}}$ where $z_1 = (z_2)^*$. Therefore, the vortex flux corresponding to $z_1$ is equivalent to an oppositely oriented vortex flux corresponding to $z_2$. Using Eq. (\ref {potential}), the static potential induced by center vortices in SU($3$) gauge group is as the following:
\begin{equation}
\label{su3}
V_r(R) =-\sum_{x} \ln\Bigl\{(1-2f_1) + 2f_1{\mathrm {Re}}\mathcal{G}_r[\alpha^{1}_C(x)]\Bigr\}.
\end{equation}
The maximum values of the angles corresponding to the Cartan generators $\mathcal{H}_{3}$ and $\mathcal{H}_{8}$ are equal to zero and $\frac{4\pi}{\sqrt{3}}$, respectively. The free parameters $f_1$, $a$ and $b$ are chosen to be $0.1$, $0.05$ and $4$, respectively. The static potentials $V_{r}(R)$ induced by center vortices for the $\{3\}$ (fundamental), $\{6\}$ and $\{8\}$ (adjoint) representations for the range $R\in [1,100]$ 
are plotted in Fig. \ref{fig7}. Figure \ref{fig8} plots the potential ratios $V_{\{8\}}(R)/V_{\{3\}}(R)$ and $V_{\{6\}}(R)/V_{\{3\}}(R)$ for contributions of center vortices.
 These potential ratios start out at the Casimir ratios: 
\begin{equation}
\frac{C_{\{8\}}}{C_{\{3\}}}=2.25, ~~~~~~~~~~~~~~~~~~~~~~~\frac{C_{\{6\}}}{C_{\{3\}}}=2.5.
\end{equation}
For the range $R \in [1,8]$, the potential ratios $V_{\{8\}}(R)/V_{\{3\}}(R)$ and  $V_{\{6\}}(R)/V_{\{3\}}(R)$ induced by the center vortices drop very slowly from Casimir ratios. Therefore, the static potentials induced by center vortices agree with the Casimir scaling.  In addition, the static potentials at large distances depend on $N$-ality which classifies the representations of a gauge group.  
 \begin{figure}[]
\centering
a)\includegraphics[width=0.450\columnwidth]{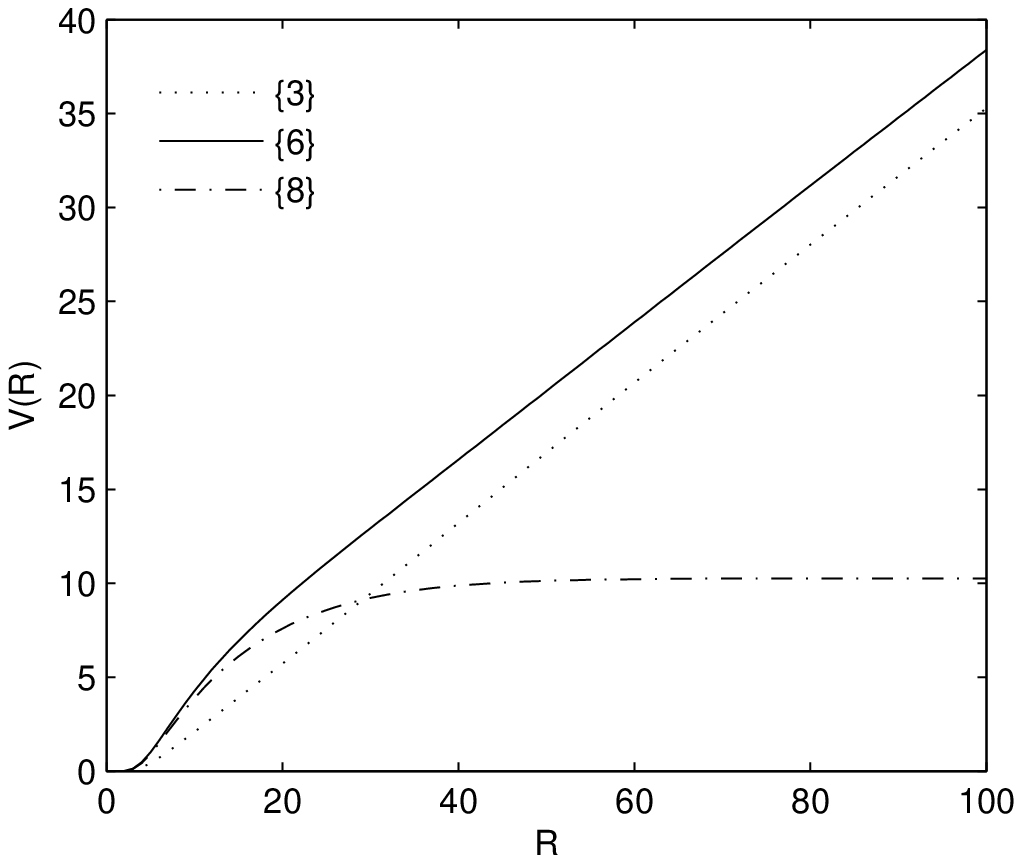}
b)\includegraphics[width=0.450\columnwidth]{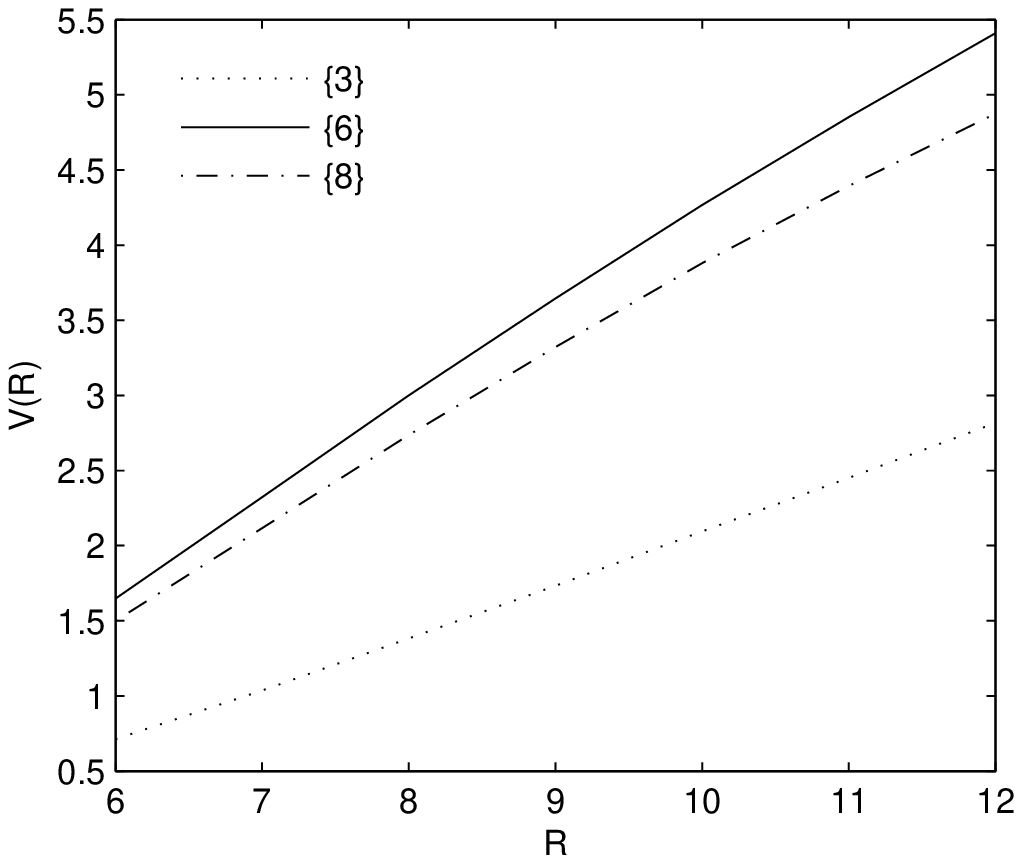}
\caption{a) The static potentials $V_{r}(R)$ induced by center vorices for the $\{3\}$ (fundamental), $\{6\}$ and $\{8\}$ (adjoint) representations for the range $R\in [1,100]$. b) same as a) but for the range $R\in [6,12]$. The potentials agree with the Casimir scaling at intermediate distances.}\label{fig7}
\end{figure}  
\begin{figure}[]
\centering
\includegraphics[width=0.450\columnwidth]{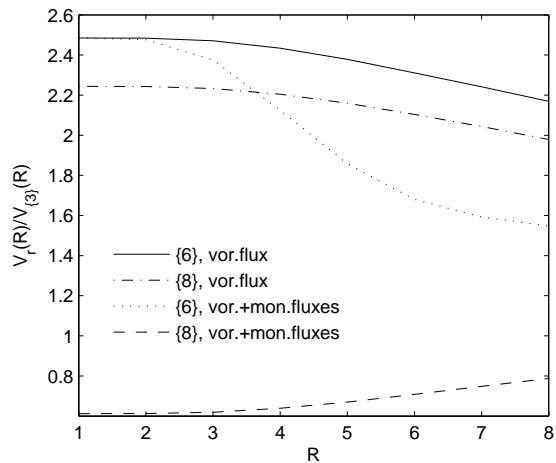}
\caption{ Ratios of $V_r(R)/V_{\{3\}}(R)$ induced by center vortices for $\{6\}$ and adjoint representations for the range $R\in [1,8]$ at intermediate distances. These potential ratios drop very slowly from Casimir ratios. Adding the contribution of monopole-antimonopole pairs to the potential ratios ruins the Casimir scaling effect. It seems that Abelian fluxes do not agree with the Casimir scaling. In the figure, `` vor. flux " means that the potential is obtained by the center vortices and also `` vor. + mon. fluxes " means that the potential is obtained by the center vortices and monopole-antimonopole pairs.}\label{fig8}
\end{figure}  
\begin{figure}[]
\centering
\includegraphics[width=0.450\columnwidth]{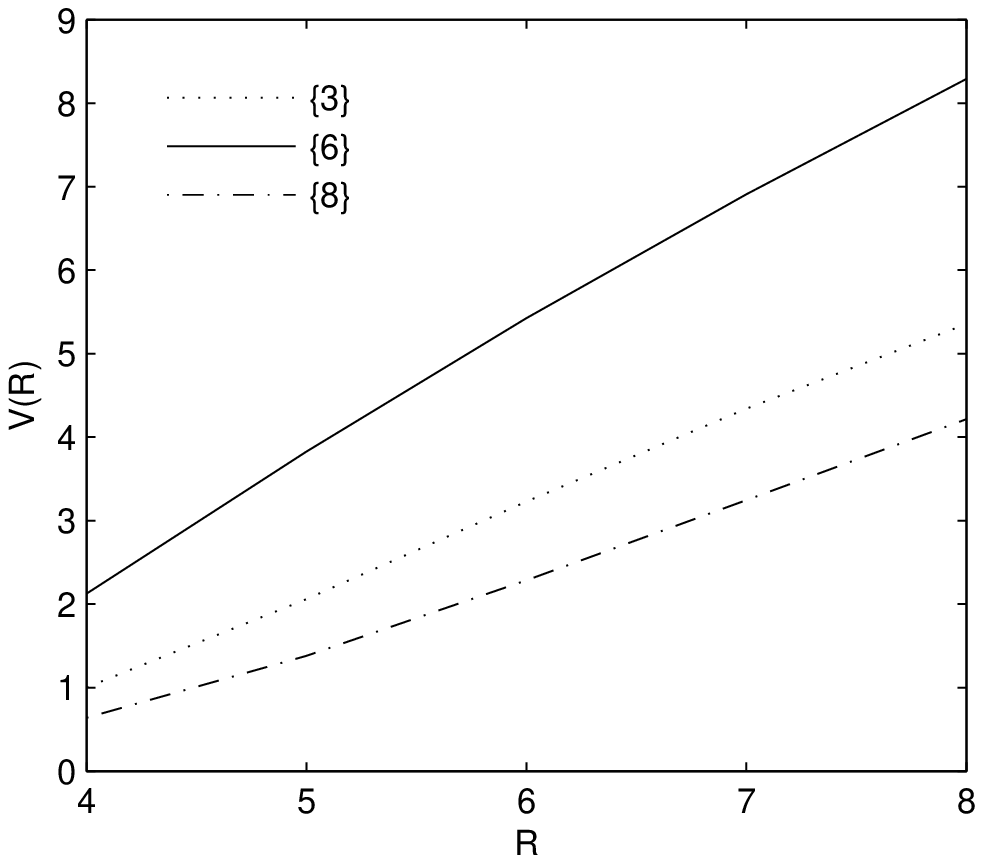}
\caption{ The contribution of monopole-antimonopole pairs is added to the static potentials of Fig. \ref{fig7} for the range $R\in [4,8]$. Adding the monopole-antimonopole pairs to the static potentials, the Casimir scaling effect is ruined at intermediate distances.}\label{fig9}
\end{figure}  

According to lattice results, monopole-vortex configurations are created in the vacuum and as we have shown in the previous section center vortices are the same objects as those line-like monopole-vortex configurations. But, in addition to monopole-vortex configurations, if we assume monopole-antimonopole pairs in the vacuum where the fluxes of three center vortices emerging from a monopole source and entering to an antimonopole are squeezed in a tube, then we would like to see whether the potentials induced by these monopole-antimonopole pairs as an Abelian fluxes agree with the Casimir scaling or not. 

Using Eq. (\ref{center}), the effect of locating a monopole-antimonopole pair, as a localized flux of three center vortex fluxes, within the Wilson loop, is as the following
\begin{equation}
\label{cen-monopole}
\exp\left[i\vec{\alpha}^{(0)}\cdot\vec{\mathcal{H}}\right]=e^{ ie(g_3-g_2)}=e^{ 2\pi i}I,
\end{equation}
where $n=0$ indicates the monopole-antimonopole pair and monopole charges satisfy the Dirac quantization condition. Therefore, the effect of the monopole-antimonopole pair on the Wilson loop is trivial ($I$) at large distances while it is non-trivial at intermediate distances. The maximum values of the angles corresponding to the Cartan generators $\mathcal{H}_{3}$ and $\mathcal{H}_{8}$ are equal to zero and ${4\pi}{\sqrt{3}}$, respectively. The contribution of the monopole-antimonopole pairs corresponding to the magnetic monopole combination of $g_3-g_2$ is 
\begin{equation}
\label{potential4}
V_r(R) = -\sum_{x}\ln\left\{ 1 -  f_{0}
[1 - {\mathrm {Re}}\mathcal{G}_r[\alpha^{0}_C(x)]\right\}.
\end{equation}
In Fig. \ref{fig9}, the contribution of the monopole-antimonopole pairs is added to the static potentials $V_{r}(R)$ induced by center vortices at intermediate distances. The free parameters $f_0$, $a$ and $b$ are chosen to be $0.1$, $0.05$ and $4$, respectively. The values of the free parameters of the monopole-antimonopole pairs are the same as those of center vortices to be able to compare the effect of these configurations on the potentials. The potential ratios $V_{\{8\}}(R)/V_{\{3\}}(R)$ and  $V_{\{6\}}(R)/V_{\{3\}}(R)$, added the contribution of the monopole-antimonopole pairs, do not agree with Casimir ratios as shown in Fig. \ref{fig8}. In other words, at intermediate distances, the potentials induced by center vortex fluxes agree with the Casimir scaling while the one induced by all fluxes of monopole do not agree with the Casimir scaling. As a result, if the vacuum contains monopole-antimonopole pairs in addition to monopole-vortex configurations, the Casimir scaling effect would be ruined. 

According to lattice results, the Abelian U($1$)$^2$ subgroup of SU($3$) cannot account for Casimir scaling, while the string  tension at asymptotic distances agree with the $N$-ality \cite{Greensite2007,Fabe1996}. Therefore, Abelian flux of monopoles destroys the Casimir scaling at intermediate distances. It seems that in Abelian theories there are Abelian flux configurations similar to monopole-antimonopole pairs which ruin the Casimir scaling effect. Since the center group $Z_3$ is the subgroup of Abelian group U($1$)$^2$, it seems there are the center vortices as well as Abelian flux of monopoles in the vacuum which lead to the $N$-ality at asymptotic distanses. Therefore, our results in the model are in agreement with the lattice gauge theory.

\section{Conclusion}
\label{sec:8}
Both the Abelian monopoles and center vortices can be condensed in the vacuum which lead to the quark confinement. Therefore, a correlation between these objects must exist. According to lattice simulations,  almost  all  monopoles  are sitting  on  top  of  center vortices. Thus, Abelian monopoles and center vortices appear to be correlated  with  each  other. Motivated by the correlations between monopoles and center vortices, in the thick center vortex model, the center vortex flux is obtained using fractional fluxes of monopoles  for SU($3$) gauge group. Combining one third of the total flux of $g_3$ monopole and one third of the total flux of $g_2$ monopole pointing in opposite direction, the center vortex flux is obtained. Since the SU($3$) monopole charges satisfy the Dirac quantization condition, two third of the total monopole flux on the Wilson loop may be regarded the same as one third of the total monopole flux pointing in opposite direction. Some configurations of the monopole fractional fluxes where combinations of them produce monopole-vortex configurations are constructed. Then the effect of the line-like configurations of the monopole fractional fluxes which are similar to center vortices is investigated in a thick center vortex-like model. Comparing the potential induced by fractional fluxes of monopoles constructing center vortex flux with the one induced by center vortices, we observe attractive energy between fractional fluxes of monopoles constructing center vortex flux. As a result, we conclude that the combination of fractional fluxes constructing the center vortex flux is stable configuration, as expected. 

On the other hand, the static potentials induced by center vortex configurations are calculated for some representations. The effect of line-like monopole-vortex configurations on the Wilson loop is the same as the effect of center vortices on the loop. The potential ratios agree with the Casimir scaling at intermediate regime. Although lattice results have
 only reported monopole-vortex configurations, but we assume monopole-antimonopole pairs in addition to the monopole-vortex configurations in the vacuum. We show that adding the contribution of the monopole-antimonopole pairs in the potentials induced by center vortices ruins the Casimir scaling at intermediate regime. According to lattice results, Abelian theories cannot account for the Casimir scaling effect at intermediate regime. It seems that in Abelian theories there are the Abelian flux configurations similar to monopole-antimonopole pairs which ruin the Casimir scaling. Therefore, only the potentials induced by the configurations with vortex fluxes appeared in lattice simulations are in agreement with the Casimir scaling effect.

\section{\boldmath Acknowledgments}

We are grateful to the Iran National Science Foundation (INSF) and the research council of the University of Tehran for
supporting this study.


\begin{thebibliography}{}

\bibitem{DelDebbio:1996mh}
 L. Del Debbio, M. Faber, J. Greensite, and {\v S}. Olejn\'{\i}k, Phys. Rev.  D \textbf{55},
  2298 (1997).
  \bibitem{Langfeld:1997jx}
 K.~Langfeld, H.~Reinhardt, and O.~Tennert, Phys. Lett.  B \textbf{419},
  317 (1998).
    \bibitem{Kovacs:1998xm}
T.~G. Kovacs and E.~T. Tomboulis, Phys. Rev.  D \textbf{57},
  4054 (1998).
      \bibitem{Alexandrou:1999iy}
C.~Alexandrou, M.~D'Elia, and P.~de~Forcrand, Nucl.Phys.Proc.Suppl. \textbf{83},
  437 (2000).
        \bibitem{Engelhardt:1999fd}
M.~Engelhardt, K.~Langfeld, H.~Reinhardt, and O.~Tennert, Phys. Rev. D \textbf{61},
  054504 (2000).
   \bibitem{Engelhardt:1999wr}
M.~Engelhardt and H.~Reinhardt, Nucl. Phys. B \textbf{585},
 591 (2000).
  \bibitem{Engelhardt:2003wm}
M.~Engelhardt, M.~Quandt, and H.~Reinhardt, Nucl. Phys. B \textbf{685},
 227 (2004).
 \bibitem{DelDebbio:1997}
{L. Del Debbio, M. Faber, J. Greensite, and {\v S}. Olejn\'{\i}k}, Center dominance, Casimir scaling, and confinement
in lattice gauge theory, In Buckow 1997, Theory of elementary particles 233-240 (1997) [arXiv:hep-lat/9802003].
\bibitem{Suzuki1993}
T.  Suzuki,  Nucl.  Phys. B, Proc.  Suppl.  {\bf 30},  176 (1993);  M.  N.  Chernodub  and  M.  I.  Po- 
     likarpov, in  Confinement, duality,  and  nonperturbative aspects of QCD, edited  by   P. 
     van  Baal  (Plenum  Press,  New  York,  1998),  p.  387 [arXiv:hep-th/9710205]; R.W.  Haymaker, 
     Phys. Rept. {\bf 315}, 153 (1999). 
\bibitem{Hooft1981}
G.~'t~Hooft, in Proceedings of the International Conference on High Energy Physics, edited by A. Zichichi (Editrice Compositori, 1976), p. 1225; S. Mandelstam, Phys. Rept.  23 C, 
     245 (1976); G.~'t~Hooft, Nucl. Phys. B190, 455 (1981). 
\bibitem{Del Debbio1998}
L. Del Debbio, M. Faber, J. Greensite, and S. Olejnik, in New Developments in Quantum Field Theory, edited by P. Damgaard and J. Jurkiewicz (Plenum Press, New York, 1998), p. 47.
\bibitem{Ambjorn2000}
 J. Ambjorn, J. Giedt, and J. Greensite, JHEP \textbf{0002},
  033 (2000). 
       \bibitem{Stack2002}
J. D. Stack, W. W. Tucker, R. J. Wensley, Nucl.  Phys. B, {\bf 639},  203 (2002).
    \bibitem{Cornwall1977}
  J. M. Cornwall, in Deeper Pathways in High-Energy Physics, Proceeding 
     of the Conference, Coral Gables, Florida, 1977, edited by B. Kursonoglu 
     et al. (Plenum, New York, 1977), p. 683.
     \bibitem{Cornwall1998}
     J. M. Cornwall,    Phys. Rev. D \textbf{58}, 105028 (1998).
          \bibitem{Chernodub2008}
    M. N. Chernodub, A. Nakamura and V. I. Zakharov, Phys. Rev. D \textbf{78}, 074021 (2008).
     \bibitem{HD2016}
S. M. Hosseini Nejad and S. Deldar, Monopole-center vortex chains in SU($2$) gauge theory, Prog. Theor. Exp. Phys. \textbf{123} B03 (2016). 
    \bibitem{Fabe1998}
 M. Faber, J. Greensite, and S. Olejnik, Phys. Rev.  D \textbf{57},
  2603 (1998).
  \bibitem{Greensite2007}
  J.~Greensite, K.~Langfeld, S.~Olejnik, H.~Reinhardt and T.~Tok,
  \emph{Phys. Rev.}  {\bf D 75} (2007) 034501.
        \bibitem{Fabe1996}
 L. Del Debbio, M. Faber, J. Greensite, and {\v S}. Olejn\'{\i}k, Phys. Rev.  D \textbf{53},
  5891 (1996).
     \bibitem {Ripka} G. Ripka, Dual superconductor models  of color conﬁnement, Lect. Notes Phys. 639 (2004) [arXiv:hep-ph/0310102].
        \bibitem{HD2015} S. M. Hosseini Nejad and S. Deldar, JHEP \textbf{09}, 039 (2015).
   \bibitem{HD2014}
 S. M. Hosseini Nejad and S. Deldar, Phys. Rev.  D \textbf{89},
  014510 (2014) [arXiv:1401.3968]. 
        \bibitem{Deldar2001}
S. Deldar, JHEP \textbf{01},
  013 (2001). 
    \bibitem{Chatterjeea2014}
C. Chatterjeea and K. Konishi, JHEP \textbf{09},
  039 (2014). 
      \bibitem{Chernodub2005}
 M. N. Chernodub, R. Feldmann, E.-M. Ilgenfritz, and A. Schiller, Phys. Rev.  D \textbf{71},
  074502 (2005).  
     \bibitem{Reinhardt}
    H. Reinhardt and M. Engelhardt, arXiv:hep-th/0010031.
  
\end{thebibliography}
\end{document}